# Coupling the Leidenfrost Effect and Elastic Deformations to Power Sustained Bouncing

Scott R. Waitukaitis[*,1,,2], Antal Zuiderwijk[1], Anton Souslov[3], Corentin Coulais[2,1], and Martin van Hecke[1,2]

[1] Huygens-Kamerlingh Onnes Laboratory, Universiteit Leiden, PO box 9504, 2300 RA Leiden, The Netherlands

[2] AMOLF, Science Park 104, 1098 XG Amsterdam, The Netherlands

[3] Instituut-Lorentz, Universiteit Leiden, PO box 9506, 2300 RA Leiden, The Netherlands

**The Leidenfrost effect occurs when an object near a hot surface vaporizes rapidly enough to lift itself up and hover[1,2]. Although well-understood for liquids[1-15] and stiff sublimable solids[16-19], nothing is known about the effect with materials whose stiffness lies between these extremes. Here we introduce a new phenomenon that occurs with vaporizable soft solids—the *elastic Leidenfrost effect*. By dropping hydrogel spheres onto hot surfaces we find that, rather than hovering, they energetically bounce several times their diameter for minutes at a time. With high-speed video during a single impact, we uncover high-frequency microscopic gap dynamics at the sphere-substrate interface. We show how these otherwise-hidden agitations constitute work cycles that harvest mechanical energy from the vapour and sustain the bouncing. Our findings unleash a powerful and widely applicable strategy for injecting mechanical energy into soft materials, with relevance to fields ranging from soft robotics and metamaterials to microfluidics and active matter.**

The Leidenfrost effect is commonly observed in the kitchen—splash a droplet of water onto a hot pan and, rather than boiling, it counterintuitively floats above the surface[1]. Far beyond a curiosity, this effect plays a critical role in industrial settings ranging from alloy production plants[4] to nuclear reactors[20] and provides a mechanism to reduce drag in fluid[4,6] and solid[19] transport. Although first described more than two centuries ago[1], issues as fundamental as droplet shape[13,14], the dynamics during impact[8,11,21], and the effects of substrate texturing[3,7,12,16,17] are only recently becoming understood. One issue that has remained unquestioned is the potential importance of the mechanical properties of the object itself. For sublimable solids such as dry ice, the Young's modulus is far too large (~10 GPa) for mechanical deformations to be relevant[16-18]. In liquids, surface tension can lead to quasi-elasticity for tiny droplets[15], but otherwise its influence is limited to capillary oscillations[2,13,14].

Here we introduce a new type of Leidenfrost effect that occurs with vaporizable soft solids—in our experiments, water-saturated hydrogel spheres (diameters 1.49±0.01 cm, masses 1.75±0.03 g). Despite consisting of ~99% water, these behave like linear elastic solids (Young's moduli $Y$=50±4 kPa; see the Supplementary Text). The effect is illustrated in Fig. 1a, where we show top-down tracks of five dyed hydrogel spheres cast onto a ceramic-coated aluminium surface at 215 °C. Immediately upon contact the spheres exhibit energetic activation, frenetically travelling around the surface at speeds of up to 0.5 m/s and emitting high-pitched screeching noises (see Supplementary Video 1). This demonstrates the potential usefulness of the effect as an energy injection strategy, particularly to create macroscopic active matter[22,23]. While the tracks convey horizontal motion, this is achieved through persistent vertical bouncing where the spheres repeatedly reach heights of 3-4 cm. The effect is long-lived—a sphere typically bounces for two to three minutes (~1000 bounces), and occasionally we observe lifetimes up to ten minutes. The activity would continue longer if the spheres were tougher—the cessation of motion is invariably associated with fracture (Fig. 1b).

With side-view video of a single sphere bouncing on a gently curved plate (see setup of Fig. 1c), we isolate the vertical motion (Fig. 1d-f and Supplementary Video 2). For a drop height of ~6 cm onto a "cold" (25 °C) surface, the sphere behaves like an inelastic ball, losing energy during each impact and quickly coming to rest (Fig. 1d). With the same drop height and a "hot" (215 °C) surface (Fig. 1e), the sphere loses energy initially, but soon reaches a steady bounce height of a few centimetres. Spheres dropped from below this height climb higher with every bounce—ultimately up to the same steady state (Fig. 1f). Simultaneous plots of the vertical trajectories and audio traces show that the screeching only occurs in the hot experiments and coincides with each impact (Fig. 2a,b).

The existence of a steady bounce height indicates that spheres impacting into a hot surface gain kinetic energy. To measure this, we first analyse the cold experiments (inset Fig. 2a) and determine the rebound curve ($H_{i+1}$ vs. $H_i$). By subtracting the anticipated "cold" rebound height from the measured one in the hot experiments, we determine the kinetic energy injected during impact with the hot surface (see Supplementary Text for full details). For steady-state bouncing, this amounts to around $10^2$ µJ (~6 mm in added bounce height), though with significant bounce-to-bounce fluctuations (inset Fig. 2b). By performing drops over a range of heights with an ensemble of similar spheres we obtain the average energy injection vs. drop height (Fig. 2c). Plotting the energy injection and loss curves on the same graph produces an intersection point at approximately 3.5 cm, *i.e.,* the steady bounce height.

One naturally suspects this behaviour is linked to vaporization of the water-saturated gel. By measuring the mass lost by spheres *vs.* how long they bounce on the hot surface (Fig. 2d) we verify this—on average, they boil ~$1.5 \times 10^2$ µg/impact. How does the vaporization process unfold?  Focusing on a single impact at significantly higher spatial and temporal resolution, we discover complex dynamics at the sphere-substrate interface.  The image sequence in Fig. 3a shows that throughout the total duration (~8 ms) of a single impact, a minute gap repeatedly opens and closes below the sphere at a much faster timescale.  This agitation is best appreciated in Supplementary Videos 3 and 4, which further reveal that each oscillation launches a Rayleigh wave that propagates around the sphere's surface.  Using the central region of the video, we see that the gap reaches heights of ~$10^2$ μm before throttling back to the surface.  The gap power spectrum (Fig. 3c) has clear peaks near 2-3 kHz.  These peaks are also present in the audio spectrum, which unveils the oscillations as the source of the audible screeching.

These dynamics are starkly different from the ordinary Leidenfrost effect, where the stable (and silent) gap is governed by a delicate balance between vaporization, viscous squeeze flow, and the object's weight[2].  In order to explain how the spheres extract mechanical energy from this vapour, we consider the factors that disrupt equilibrium. First, recent experiments with impinging liquid droplets show that for sufficiently high impact velocities the vapour layer is squeezed out entirely[11,21]. In that situation, the enhanced thermal conductivity during contact leads to accelerated vaporization and a barrage of bubbles that tear upwards through the liquid. For our impacting spheres, we also expect physical contact and accelerated vaporization, but the integrity of the solid gel prevents vapour from escaping—a trap forms.  Second, whereas liquids store no elastic energy, and stiff solids like dry ice barely deform at all, the spheres in our experiments are solid yet *soft*, which means that pressurized vapour can be converted into mechanical energy through elastic deformation.

Based on these observations, we now propose a picture for the underlying physics that recasts each gap oscillation as a thermodynamic cycle that does mechanical work on the sphere (Fig. 4a). The first stage of the cycle commences when the sphere comes into physical contact with the surface. This causes rapid vaporization, but the vapour is trapped and consequently pressure builds up below the sphere. The volume of this trap grows until until the radius reaches the Hertzian contact radius and the height reaches some value $l^*$. Now the gap opens and stage two begins where the vapour is blown out by the overpressure. After the gap reaches its maximum height, stage three begins and the sphere bottom elastically recoils toward the surface, thus reinitiating stage one. The asymmetry of the pressure evolution on the upward/downward strokes of this cycle renders the area enclosed in the pressure-volume (PV) diagram greater than zero, which results in an increase in the sphere's mechanical energy. Remarkably, this energy injection is achieved with the fuel (water), mechanism (gap oscillations) and mechanical output (increased kinetic energy) embedded in a single soft material—the sphere is effectively a *soft engine* that harvests energy from the hot surface.

Quantifying this process from first principles involves a complex interplay of vaporization, gas dynamics, and deformation. We now put forth a simplified numerical model that couples these three essential ingredients. As illustrated in Fig. 4b, we mimic the soft sphere with a one-dimensional chain of *N* identical masses (mass *m*) connected by *N*-1 identical springs (rest length *δ*, stiffness *κ*). By specifying the forces and initial conditions, we numerically integrate the equations of motion and solve for the dynamics. The hard surface is modelled by a stiff spring that acts on the bottommost mass once it passes *y*=0. Impacts onto a "hot" surface include an additional force that arises from the vapour pressure, which we approximate as linear growth during the build-up ($\dot{P} = \alpha$) and exponential decay during the escape ($\dot{P} = -P/\tau$). The pressure acts over an area, *A*, that evolves throughout impact according to the Hertzian overlap of the sphere with the surface. As in the PV diagram of Fig. 4a, we define the build-up stage as beginning each time the bottommost mass reaches the surface and continues until it rises above the maximum trap height before the gap opens, *l\**. Full details for the simulations can be found in the Supplementary Text.

We estimate the parameters for our model based on our experimental data (values are stated in the caption to Fig. 4 and the full calculations are explained in the Supplementary Text). Despite its simplicity, the model qualitatively and semi-quantitatively captures all of the experimental observations (for qualitative comparison, see Supplementary Video 5). During impact, the position of the lowest mass in the chain, *i.e.,* the gap, rises up to heights on the order of $10^2$ μm with a frequency around ~2.5 kHz, thus reproducing the gap oscillations (Fig. 4c). Calculating the kinetic energy injection and loss curves exactly as in the experiments, we see an intersection point at drop heights of a few centimetres. Crucially, this illustrates how our simple formulation of the the coupling between vapour release and the elastic deformations of the soft sphere leads to macroscopic energy harvesting and therefore persistent bouncing (Fig. 4d).

We have introduced a new type of Leidenfrost effect that occurs with vaporizable soft solids—rather than hovering, they exhibit energetic activation in the form of persistent bouncing. Our experiments and minimal model reveal that the mechanism is the coupling between vapour release and elastic deformations, which lead to microscopic work cycles at the sphere-substrate interface that inject mechanical energy. Beyond this fundamental result, our findings provide a robust and immediately applicable strategy for activating and actuating soft systems in general. Hydrogels are already widely employed owing to their desirable elastic and electrical properties, biocompatibility, and bondability to other materials[24-27]. Working with solid objects opens up the tantalizing possibility to change the dynamics with an object's shape, and pre-existing technologies for hydrogel fabrication with droplet detachment[28], molding[29], and direct 3D printing[30] puts scalable production for large systems within reach. The elastic Leidenfrost effect therefore has direct ramifications for studies in soft robotics, microfluidics, metamaterials, and active matter, where activation and actuation of soft materials play a critical role.

**Acknowledgements**

We acknowledge K. Harth for a critical reading of the manuscript. We thank E. Clement, J. Dijksman, D. Durian, H. Jaeger, D. Lohse, and L. Lubbers for productive conversations. M. van Deen, D. Ursem, R. Struik and J. Mesman provided critical technical support. A.S. acknowledges funding from the Delta Institute for Theoretical Physics and the hospitality of the IBS Center for Theoretical Physics of Complex Systems, Daejeon, South Korea. We acknowledge funding from the Netherlands Organisation for Scientific Research through grants VICI No. NWO-680-47-609 (M.v.H. and S.R.W.) VENI No. NWO-680-47-445 (C.C.), and VENI No. NWO-680-47-453 (S.R.W.).

**Author Contributions**

S.R.W. conceived of the project. S.R.W. and A.Z. performed the experiments. S.R.W., M.v.H., C.C., and A.S. developed the model and S.R.W. implemented it numerically. All authors contributed to the writing of the manuscript.

**Competing Financial Interests**

The authors declare no competing financial interests.

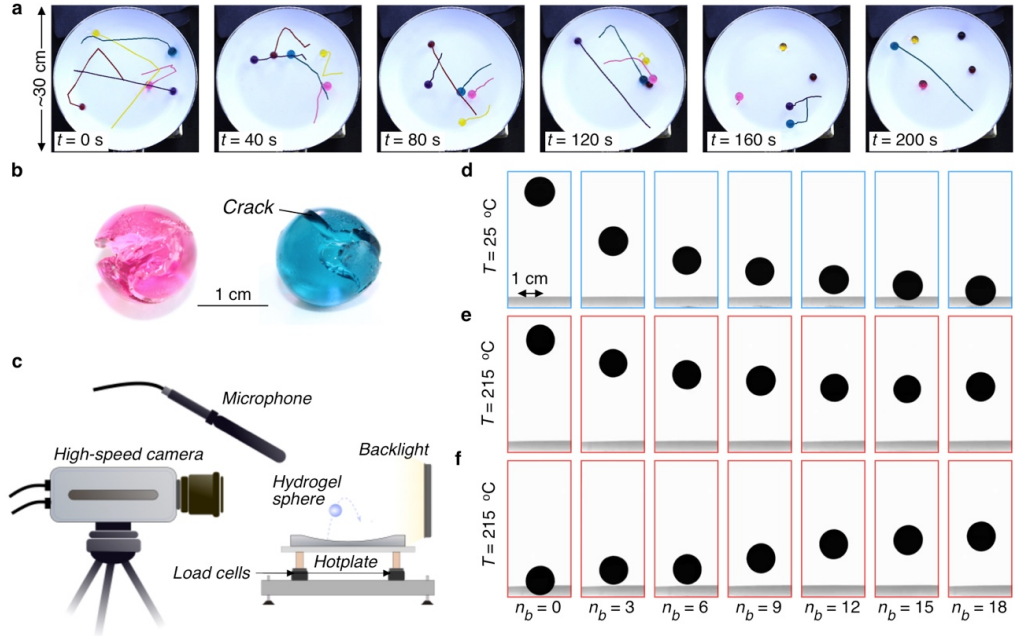

Figure 1: Persistent bouncing of hydrogel spheres on a hot surface. **a** Top-down stills showing the long-lasting dynamics of five hydrogel spheres dropped onto a hot (215 °C) surface. The lateral motion is mediated by vertical bouncing, and the spheres emit high-pitched screeching noises throughout (see Supplementary Video 1). Lines show tracks of the preceding 0.42 s. **b** Spheres typically stop after 2-3 minutes as a result of fracture. **c** Main experimental setup with a high-speed camera, backlighting, a microphone, and dynamic load cells to determine contact intervals. **d-f** Side-view stills showing the maximum height for bounce number $n_b$ with spheres dropped from **d** ~6 cm onto a "cold" (25 °C) surface and **e** "hot" (215 °C) surface, and from **f** ~2 mm onto a hot surface (see Supplementary Video 2). The sphere dropped onto the cold surface comes to rest, while the spheres dropped onto the hot surface reach a steady bounce height of about 3.5 cm.

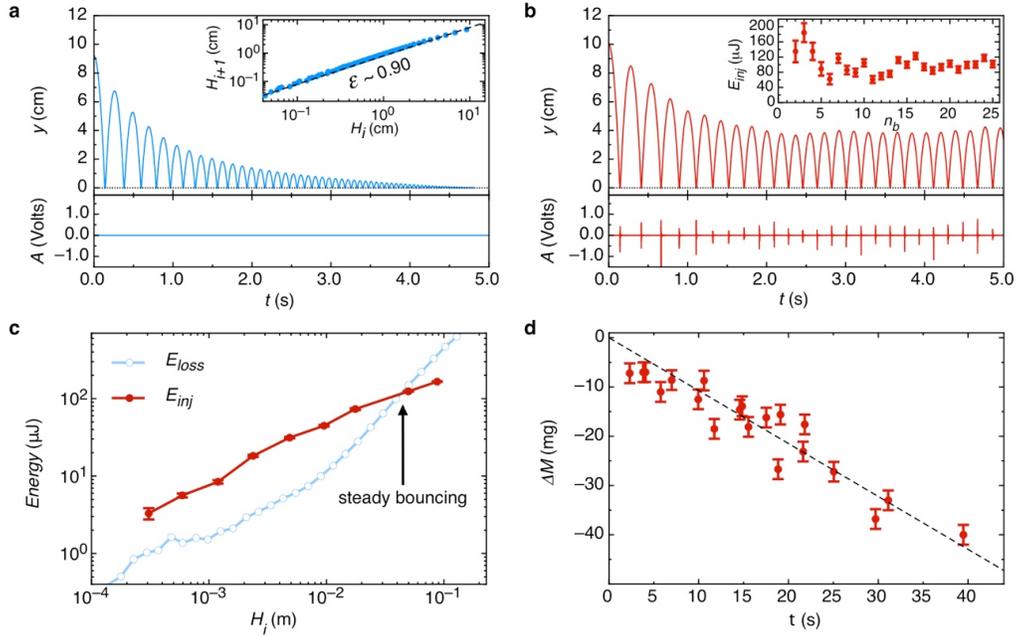

Figure 2: Energy injection and mass loss. **a** Vertical position (top) and audio (bottom) *vs.* time for a sphere bouncing on a cold surface. The rebound curve (inset) reveals the restitution coefficient, $\varepsilon$, is nearly constant (see Supplementary Text). **b** Same as (a) for a sphere on a hot surface. The audio trace reveals that screeching only occurs during impact. The inset shows the kinetic energy gained, $E_{inj}$, during each impact on the hot surface (see Supplementary Text). **c** The kinetic energy lost during impacts on a cold surface (blue open circles) and injected during impacts on a hot surface (red closed circles) *vs.* drop height. **d** The mass lost *vs.* time is 1.0±0.2 mg/s (dashed line) or about $1.5 \times 10^2$ μg/impact.

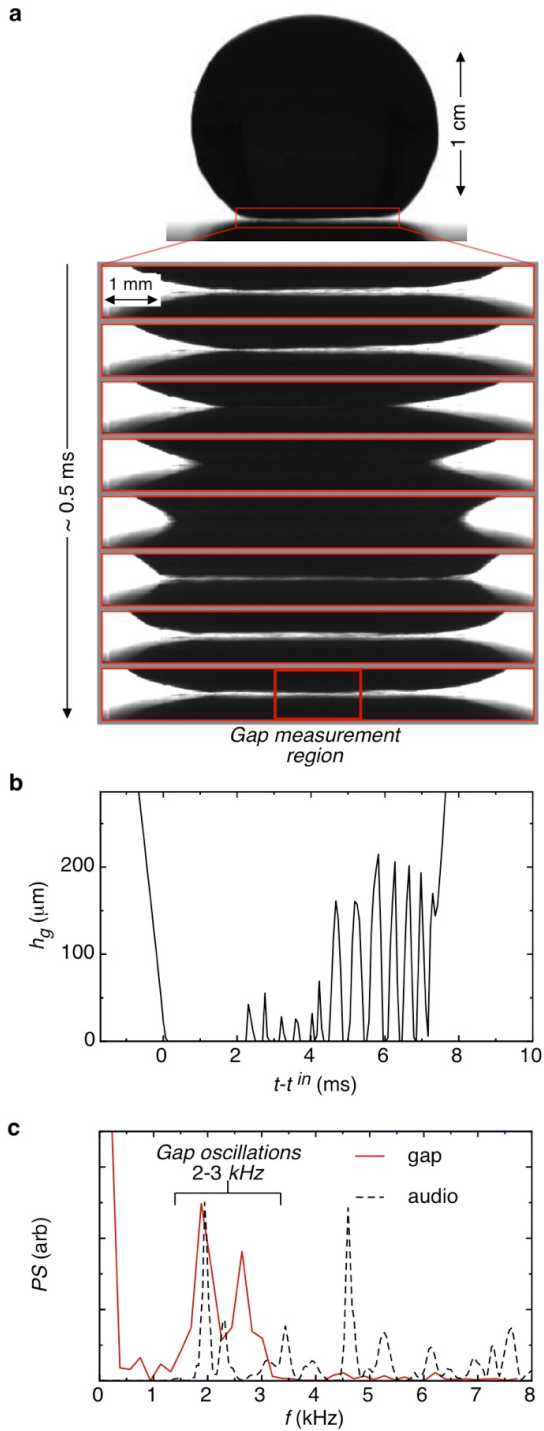

Figure 3: High-frequency, microscopic gap oscillations at the interface **a** High-speed video stills of a single impact ($H_i \approx 3.5$ cm) at high magnification and frame rate (15625 fps) reveal that a minute gap below the sphere rapidly opens and closes many times during each impact (see Supplementary Videos 3 and 4). The timescale for one cycle is about 0.5 ms. **b** Plotting the gap height (averaged over the central 100 pixels) *vs.* time shows that the deformation of the sphere underbelly is on the order of $10^2$ μm. **c** The power spectra of the gap and the audio signal both have clear peaks around 2-3 kHz, which indicates that the gap oscillations are the source of the screeching.

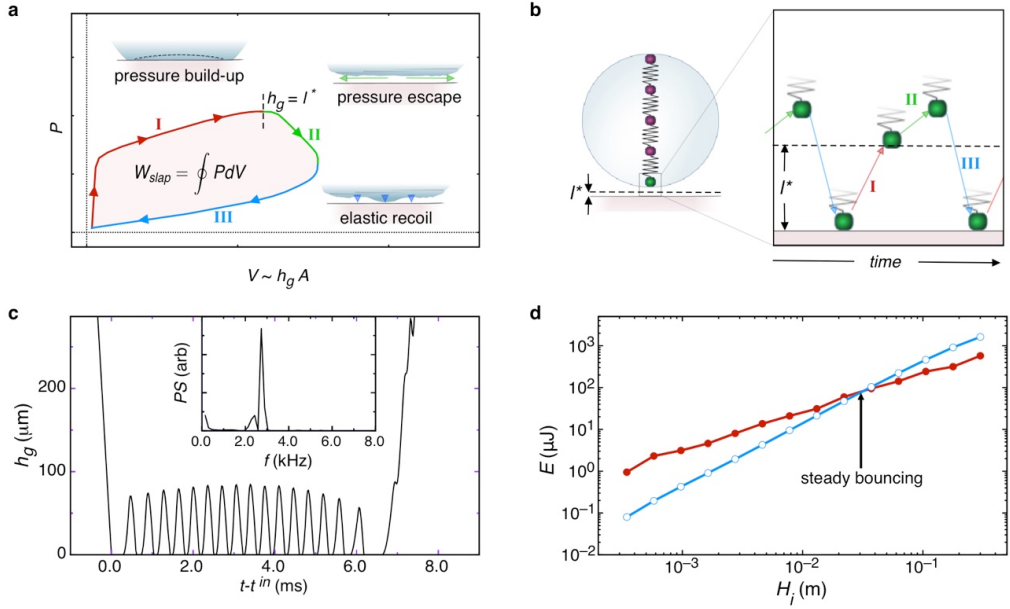

Figure 4: Coupling the Leidenfrost effect to elastic deformations. **a** Sketch and PV diagram for the gap oscillation work cycle consisting of three stages: pressure build-up (I), pressure escape (II), and elastic recoil (III). **b** We simulate the sphere as a chain of $N$ identical point-masses (mass $m$) connected by $N$-$1$ identical neo-Hookean springs (stiffness $\kappa$, rest length $\delta$). The pressure grows linearly ($\dot{P} = \alpha$) after each time the lowest mass reaches the surface (stage I) and then decays exponentially ($\dot{P} = -P/\tau$) after each time the lowest mass rises above the trapping height $l^*$ (stage II and III). **c-d** Simulation results for $N$=201 masses and all other parameters estimated from our experiments ($m$=8.7x10$^{-5}$ kg, $\delta$=7.5x10$^{-5}$ m, $\kappa$=6.0x10$^{-1}$ N, $l^*$=50 μm, $\alpha$=1.5x10$^8$ Pa/s and $\tau$=5.0x10$^{-5}$ s—see Supplementary Text). **c** The lowest mass, *i.e.* the gap, opens and closes to heights on the order of $10^2$ μm at ~2.5 kHz (power spectrum in inset). **d** Plots of the kinetic energy injected and lost *vs.* drop height for the model shows the same steady-state bouncing phenomenology as the experiments with a similar equilibrium bounce height of a few centimetres. For rendered videos of the model gap oscillations and steady-state bouncing, see Supplementary Video 5.

Supplementary Materials for:

# Coupling the Leidenfrost Effect and Elastic Deformations to Power Sustained Bouncing

Scott R. Waitukaitis[*,1,,2], Antal Zuiderwijk[1], Anton Souslov[3], Corentin Coulais[2,1], and Martin van Hecke[1,2]

[1] *Huygens-Kamerlingh Onnes Laboratory, Universiteit Leiden, PO box 9504, 2300 RA Leiden, The Netherlands*

[2] *AMOLF, Science Park 104, 1098 XG Amsterdam, The Netherlands*

[3] *Instituut-Lorentz, Universiteit Leiden, PO box 9506, 2300 RA Leiden, The Netherlands*

**Introduction** This document contains Supplementary Materials for the main text in the following order: (1) Supplementary Text for the main manuscript, which details the the methods for sample preparation and measurements and an in depth discussion of the numerical model, (2) Supplementary Figures 1-6, and (3) the Descriptions for the Supplementary Videos.

**Sphere preparation** We prepared commercially available hydrogel spheres (Educational Innovations Inc. ® GB-710) by adding dehydrated specimens to a mildly saline solution (0.6 g NaCl/KCl table salt per 1.0 L Milli-Q® water). As shown in Supplementary Fig. 1, the distribution (mean and spread) of masses of the dehydrated spheres is 24±1 mg, while for the swollen spheres it is 1.75±0.13 g. The water content by mass is therefore 98.6±0.1%. Given that the Young's moduli change quickly with sphere size (Supplementary Fig. 2), we performed experiments with a subset of spheres that had a distribution $M=1.75\pm0.03$ g. For imaging data, we dyed the otherwise clear spheres with food colouring (Rainbow Dust ProGel®).

**Surfaces**  For the data in Figs. 1d-f, Figs. 2a,b and the energy loss measurements in Fig. 2c, we used an aluminium surface with a gentle spherical curvature (radius 81.9 cm) to keep the sphere within the field of view.  In cold experiments we applied a superhydrophobic coating (Glaco® Mirror Coat Zero) to mitigate wetting.  Vaporisation prevented wetting in hot experiments.  The aluminium surface permitted observation of sequential impacts, but it easily became sullied.  This made subsequent interactions erratic and required constant cool-down so it could be cleaned. For the energy injection measurements of Fig. 2c, we used a flat, ceramic-coated aluminium surface.  This permitted us to observe only a few bounces at a time, but allowed us to clean the surface while hot and avoid cool-down.  The roughness of both the flat and curved surfaces was less than 5 μm.  We heated the surfaces with a hot plate (Stuart US150 Hot Stirrer, 700 W) and measured their temperatures to within ±5 °C with a thermocouple.

**Sphere Young's modulus measurements**   We characterized the spheres' mechanical properties using an Instron (model 5965) equipped with a 10N load cell to take force-displacement curves for individual specimens sandwiched between two vertical crossheads (Supplementary Fig. 2a). We attached fine (1200 grit) sandpaper to the crossheads to prevent slippage.  We coated the sandpaper with superhydrophobic spray (Glaco® Mirror Coat Zero) to mitigate wetting. The Young's modulus was calculated by fitting the force-displacement curve to the equation for a Hertzian sphere compressed between two hard half-spaces[35], *i.e.*,

$$F = \frac{Y\sqrt{d}}{3(1-v^2)} \Delta^{3/2}. \qquad (1)$$

Here *Y* is the Young's modulus, *d* is the diameter, *v* is the Poisson's ratio (*v*=0.5), and *Δ* is the crosshead displacement. The value of the Young's modulus varies from sphere to sphere and with the diameter (Supplementary Fig. 2b). For the spheres we used in the experiments, the distribution of Young's moduli is *Y*=50±4 kPa.

**Sphere trajectory analysis**  Our experimental setup provided a variety of ways to measure a sphere's vertical trajectory. While the most straightforward would seem to be with the camera, this has the disadvantages of (1) poorly resolving small drop heights and (2) requiring inconveniently large amounts of data and analysis. Instead, we used the force sensors to define the contact intervals and backed out the vertical trajectories from Newton's laws[36-39] (Supplementary Fig. 3a). We take $t_i^{out}$ to be the time when impact *i* ends and $t_{i+1}^{in}$ to be the time when impact *i*+1 begins and furthermore define $\Delta t = t_{i+1}^{in} - t_i^{out}$ and $\bar{t} = (t_{i+1}^{in} + t_i^{out})/2$. The maximum height in the parabolic flight between is $H_{i+1} = g\Delta t^2/8$, where *g*=9.8 m/s². The trajectory is given by

$$y(t) = H_{i+1} - \frac{1}{2}g(t - \bar{t})^2. \qquad (2)$$

This is valid as long as aerodynamic drag is small compared to the sphere's weight. For spheres with a diameter of 1.5 cm moving through air at 1.0 m/s (the maximum velocity we encounter), the drag term is on the order of $\rho_{air}\pi d^2 V^2/4 \sim 1\times10^{-4}$ N—two orders of magnitude smaller than the weight.

For impacts on the hot surface we modified this procedure on account of the gap oscillations, which modulate the force sensor signal at high frequency. Additionally, the sensor picks up other spurious oscillations ranging from 1-10 kHz due to mechanical resonances in our setup. We therefore quantified the response of an impact onto a hot surface by first performing a low pass (<1 kHz) filter and then defining $t_c$ and $F_{max}$ the same way as in the cold experiments (Supplementary Fig. 3b). The global features of impact involve low enough frequencies (<500 Hz) to be preserved.

**Energy injection** We measured the kinetic energy injection by comparing the rebound heights for spheres dropped on cold and hot surfaces. The bouncing on a cold surface is close to what would be expected for a constant coefficient of restitution, which would give $H_{i+1} = \varepsilon^2 H_i$. However, $\varepsilon$ deviates from constant behaviour at low and especially at high $H_i$ (Supplementary Fig. 4). To account for this, we binned our data and constructed an interpolated coefficient of restitution curve for the cold surface, $\varepsilon(H_i)$, as shown in Supplementary Fig. 4b. The energy injection is then given by

$$E_{inj}(H_i) = Mg(H_{i+1} - \{\varepsilon(H_i)\}^2 H_i). \qquad (3)$$

We have verified that the form of the curve, $\varepsilon(H_i)$, is not consistent with dissipation from plastic deformation[36] or viscoelasticity[40]. This reveals that the energy lost during impact is mainly transferred to spheroidal oscillations[41], which are clearly visible in Supplementary Video 3. These oscillations damp out over the long parabolic flights between impacts. Note these spheroidal oscillations are large wavelength, *i.e.*, $\lambda \sim d$, and are not the same as the short-wavelength Rayleigh waves ($\lambda \sim 1$mm) that are a result of the gap oscillations (visible only in the second part of Supplementary Video 3).

**Separation of impact and gap oscillation length and time scales** Our arguments implicitly rely on the assumption that the overall Hertzian impact and the gap oscillations occur independently. We now verify this assumption. Hertzian theory[36] predicts how the maximum force, $F_{max}$, and total contact time, $t_c$, should scale with the drop height:

$$F_{max} = k_1^{2/5} \left(\frac{5}{2} M g H_i\right)^{3/5}, \text{ and} \quad (4)$$

$$t_c \approx 3 \left(\frac{M}{k_1}\right)^{2/5} (g H_i)^{-1/10}. \quad (5)$$

Here, $k_1 = 4Y\sqrt{d/2}/3(1-\nu^2)$. Using our other measurements, we predict the trends for $t_c$ and $F_{max}$ *vs*. $H_i$ for all impacts with no free parameters (Supplementary Figs. 3c,d). The deviations at lower $H_i$ are consistent with the increasing importance of gravity for small drop heights[36]. Given that (1) the data for $t_c$ and $F_{max}$ for hot/cold impacts lie on top of each other and (2) follow the trends predicted by Eqs. 4 and 5, we confirm that the gap oscillations can be seen as a perturbation to Hertzian impact.

**Numerical model: equations of motion** In our numerical model, we solve for the dynamics of a one-dimensional chain of alternating masses and springs as it bounces off of a hard surface. The chain consists of $N$ identical masses connected by $N$-1 identical springs in series. Given the initial positions and velocities for each mass and the forces that act on them, we numerically integrate the equations of motion to solve for the dynamics. Denoting the index of the bottom mass as $i=0$ and counting upwards, these are

$$m\ddot{y}_0 = -f_0 + f_s + PA - gm$$
$$m\ddot{y}_i = (f_{i-1} - f_i) - gm \quad i \neq 0, N-1 \quad (6)$$
$$m\ddot{y}_{N-1} = f_{N-2} - gm$$

Here, $m$ is the (common) mass of each object in the chain, $g$ is the acceleration due to gravity, $f_i$ is the compressive force in the spring between masses $i$ and $i+1$, $f_s$ is the force via interaction with the hard surface, $P$ is the pressure provided by the vapour and $A$ is the instantaneous contact area.

**Parameters of the spring-mass chain** For $f_i$, we use neo-Hookean springs to prevent the masses from passing through each other[42]. We denote the positions of the masses by $y_i$ and the (common) rest length of the springs as $\delta$. Defining the stretch of the $i^{\text{th}}$ spring as $\lambda_i = (y_{i+1} - y_i)/\delta$ and the (common) spring constant as $\kappa$, then the force between the masses $i$ and $i+1$ is given by $F_i = \kappa(\lambda_i - \lambda_i^{-2})$. In the limit of small deformations ($\lambda_i \sim 1$), these springs are approximately Hookean with stiffness $3\kappa/\delta$. As discussed previously, the spheres lose their centre-of-mass kinetic energy during impact to excitation of large-wavelength spheroidal modes. Analogously, the energy lost for the spring-mass chain arises from excitation of large-wavelength longitudinal modes. For the data in Fig. 4d, we simulate one bounce at a time and do not consider any damping that occurs in flight between bounces. This is also true for Supplementary Video 5, where we simulate individual bounces and then extrapolate the parabolic flights in between.

We constrain the sum of the individual masses to equal the sphere mass ($Nm=M$), and the sum of the spring rest lengths to equal the sphere diameter [ $(N-1)\delta=d$ ].  For the neo-Hookean spring constant, $\kappa$, we compare numerical and experimental results for impacts onto a "cold" surface and pick the value that minimizes the deviation of the *average* force throughout impact *vs.* drop height.  The motivation for this procedure is illustrated in Supplementary Fig. 5a, where the smooth $F$ *vs.* $t$ curve for an experiment is plotted alongside the step-like curve from a chain.  The step arises because the chain impact is dominated by a shock that gives a nearly constant force $F(t) \sim cV_0\sigma$, where $c=\sqrt{3\kappa\delta/m}$ is the sound speed (~4 m/s), $V_0$ the impact velocity, and $\sigma=M/d$ the linear mass density.  The step shape suggests a convenient strategy for mimicking the sphere.  Namely, for an appropriate value of $\kappa$, the average forces can nearly match up (Supplementary Fig. 5b).  To find this value, we first run impacts at a fixed value of $\kappa$ for several different drop heights (spread out over our experimental range of approximately 400 μm to 20 cm) and calculate the sum of the squared residuals between the average force and the prediction from Hertz.  Calculating this quantity for different values of $\kappa$ reveals a clear optimum (Supplementary Fig. 5c), which produces an $F_{av}$ *vs.* $H_i$ curve that nicely approximates the experimental results (Supplementary Fig. 5b).  Advantageously, this procedure ensures that the contact times for the experiments and simulations are similar (Supplementary Fig. 5d).  For the data presented in Fig. 4 we use $N=201$, $m=8.7 \times 10^{-5}$ kg, $\delta=7.5 \times 10^{-5}$ m, and $\kappa=6.0 \times 10^{-1}$ N.  The results start becoming independent of discretization for $N$ larger than 50 (see Supplementary Fig. 6).

**Interaction with the hard surface**  For the interaction with the hard surface, $f_s$, we use the penalty method[43] and turn on a very stiff spring for the bottom mass once it passes below $y=0$. Additionally, we use the FEM practice[43] of incorporating a damping term on the bottom-most mass when it is below $y=0$ to stabilize the contact. Concretely, this force is $f_s = (-k_s y_0 - \beta_s \dot{y}_0)\,\theta(-y_0)$. Here $k_s$ is a spring constant, $\beta_s$ is a damping coefficient, and $\theta(-y_0)$ is the Heaviside function to reflect that this only turns on for $y_0<0$. In order for the surface to be "hard," it must be the case that $k_s >> 3\kappa/\delta$. The damping parameter, $\beta_s$, should be just large enough to maintain contact each time the lowest mass passes zero (until it is pushed up again by the growing pressure). Beyond these criteria, the exact values of $k_s$ and $\beta_s$ do not significantly alter the gross features of impact, although extremely large values unnecessarily slow down computation. For the data in the main text we use $k_s = 10^5 \kappa/\delta$ and $\beta_s = 500$ Ns/m.

**Evolving contact area**  We account for the evolving contact area of an impacting sphere in each step of our numerics as is done in Hertzian contact mechanics. Explicitly, we calculate the intersection area of a sphere of diameter $d$ located at the centre-of-mass of the chain, $Y_{cm}$, with the plane located at $y=0$. This is given by $A=\pi[(d/2)^2-(d/2-Y_{cm})^2]\,\theta(d/2-Y_{cm})$. The Heaviside function, $\theta(d/2-Y_{cm})$, is present to satisfy the requirement that the contact area is zero if $Y_{cm} > d/2$. Having an evolving contact area is not necessary for energy injection. However, an evolving contact (1) better approximates the experimental situation where the energy injected per cycle is larger near the middle of each impact than at the beginning or end and (2) produces a smooth curve by rendering the effect of the discrete total number of cycles less pronounced.

**Pressure evolution** The vapour does work on the sphere to increase its mechanical energy—this implies that the pressure evolution during the upward and downward branches of each gap cycle must be different. This type of asymmetry arises naturally from vapour trapping. To produce trapping in our 1D model, we initialize the pressure to $P=0$ Pa and use the following pressure evolution:

$$\dot{P} = -P/\tau \qquad \text{until } y_0 \text{ reaches } 0$$
$$\dot{P} = \alpha \qquad \text{until } y_0 \text{ reaches } l^* \qquad (7)$$
$$\dot{P} = -P/\tau \qquad \text{until } y_0 \text{ reaches } 0$$
$$\vdots$$

Here, $\tau$ is a decay timescale, $\alpha$ is a pressure growth rate, and $l^*$ is the gap height at max pressure. To achieve semi-quantitative agreement with our model, we now estimate the appropriate scale for the values of $l^*$, $\alpha$, and $\tau$ based on our experimental data.

First, we guess a reasonable value for the length scale, $l^*$, which is limited to be smaller than, but on the order of, the gap height. Realistically, it will change for each cycle of a single impact owing to (a) the changing pre-compression of the sphere and (b) the changing contact area. It will also change from one impact to the next owing to differences in the dynamics as a function of the drop height. Practically, however, we use the zeroth order approximation of a constant value because it is sufficient to recover the observed behaviours. We choose our estimate for the constant value based on observations of an equilibrium bounce and use $l^*= 50$ μm—on the order of maximum gap height seen in the middle of an impact for a sphere dropped from the equilibrium bounce height (Fig. 3b).

To estimate the value for the pressure build-up rate, $\alpha$, we are guided by two calculations. First, as is evident in Fig. 3B, the gap under the sphere remains closed in the middle of the impact for $\Delta t \approx 0.25$ ms. When the contact fully breaks, the upward force provided by the trapped vapour must be at least as large as the downward force from the Hertzian compression above. For a drop height of 3.5 cm, the Hertzian force from the compressed sphere above in the middle of the impact, $F_{max}$, is on the order of 1 N (Fig. S3C). With the aid of Hertzian theory or from our experimental data (*e.g.*, Supplementary Videos 3 and 4), we also know that the Hertzian contact area is $A_{max} \approx 10^{-4}$ m$^2$. This gives the lower bound for the estimate, $\alpha \approx F_{max}/A_{max}\, \Delta t \approx 4 \times 10^7$ Pa s$^{-1}$.

As a second independent calculation, we use the data for the mass loss in combination with the ideal gas law. Again considering a drop from the steady bounce height, the mass lost is 150 µg/impact (Fig. 2d). Assuming 10 gap oscillations per impact and additional evaporative losses when the trap is open, we roughly estimate that the amount of vapour trapped during one cycle is on the order of 10 µg, which amounts to $n = 6 \times 10^{-7}$ mol. This is contained in a volume that scales like $A_{max}\, l^*$. The time it takes for this to develop is, once again, $\Delta t \approx 0.25$ ms. Using the ideal gas law, we therefore have $\alpha = nRT/A_{max}\, l^* \Delta t \approx 2 \times 10^8$ Pa s$^{-1}$ ($R$ is the universal gas constant and $T \approx 500$ K the temperature of the surface). This value is higher than our other estimate, and we expect this arises because the trapping process is not perfect. This is evidenced by the initially choppy development of the gap oscillations (Supplementary Video 4). For the simulation data in Fig. 4, we use a value between these estimates, $\alpha = 1.5 \times 10^8$ Pa s$^{-1}$.

The decay term in the pressure evolution reflects the fact that once the gap opens up the trapped vapour can escape. We use exponential decay to recover this behaviour. To inform our decision on a reasonable value for the time constant, $\tau$, we consider the situation of pressure driven evacuation of viscous vapour from a fixed gap (height $l^*$) trapped between two flat disks (area $A_{max}$). Accounting for Poiseuille's flow and mass conservation, one can show that this system obeys Darcy's law[17,44] with a timescale given by $\tau = 12\eta A_{max}/\pi l^{*2} \Delta P$, where, $\eta$ is the viscosity of the vapour ($2 \times 10^{-5}$ Pa s) and $\Delta P$ is the pressure difference between the center and the edge ($\Delta P \approx F_{max}/A_{max} \approx 10$ kPa). This gives a timescale of $10^{-4}$ s, which we consider an upper bound for two reasons. First, the gap opens to heights greater than $l^*$, which further reduces the escape time. Second, the increasing volume of the gap itself reduces the pressure via the ideal gas law. In the simulations we use a slightly smaller value of $\tau = 5 \times 10^{-5}$ s. Our model is not terribly sensitive to this parameter so long as it is not significantly larger than our upper bound—energy injection still occurs for infinitesimal values of $\tau$ as long as the lengthscale, $l^*$, and the pressure buildup parameter, $\alpha$, are greater than zero.

**Supplementary Figures**

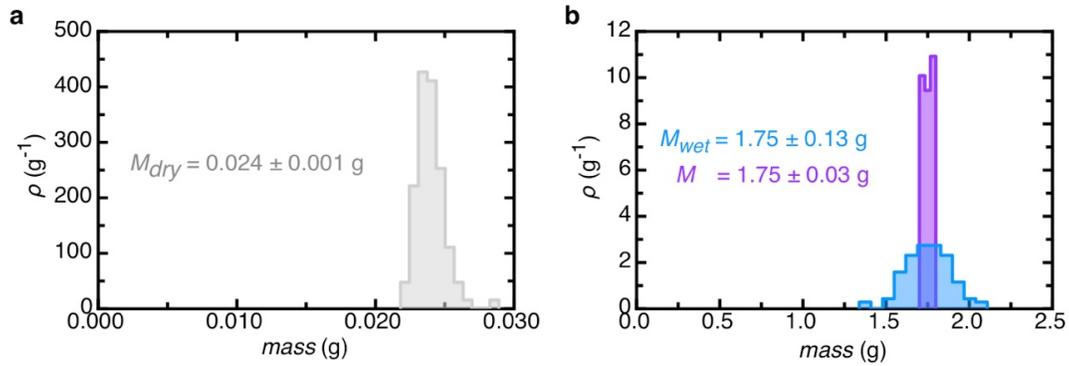

Supplementary Figure 1: Hydrogel sphere preparation and water content. **a** Plot of the mass distribution of dry hydrogel spheres with mean and spread $M_{dry}$=2.4±0.1 mg. **b** Plots of the mass distributions of all water-saturated spheres (blue, $M_{wet}$=1.75±0.13 g) and the subset of spheres used in the experiments (purple, $M$=1.75±0.03 g). We used the subset because the Young's modulus changes significantly with the sphere size (Supplementary Fig. 2). Using the measurements for the average wet and dry mass, the water content of our spheres is 98.6±0.1%.

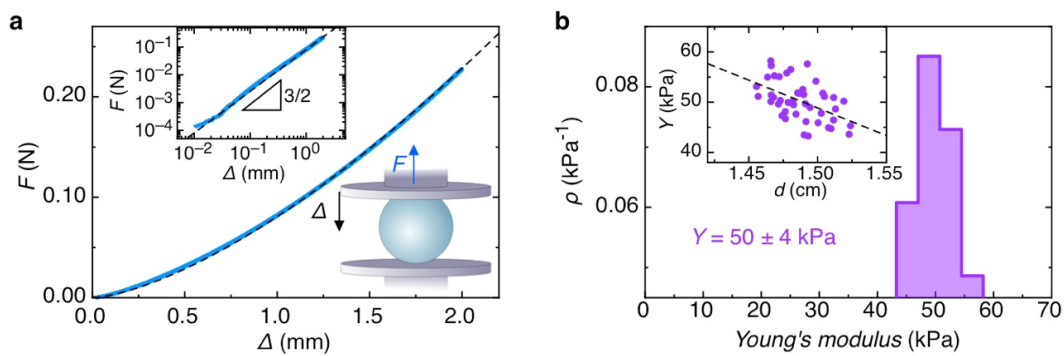

Supplementary Figure 2: Mechanical characterization. **a** Force-displacement curve for a swollen hydrogel sphere compressed between two vertical crossheads. The curve follows a 3/2 power law consistent with Hertz's theory for a sphere made from a linear, isotropic material. By fitting such a curve to Eq. 1, we extract the Young's modulus. **b** The distribution of Young's moduli for the subset of spheres used in the experiments has a mean and spread of 50±4 kPa. As the inset shows, the moduli change with the sphere size—this encouraged us to use a tight size distribution.

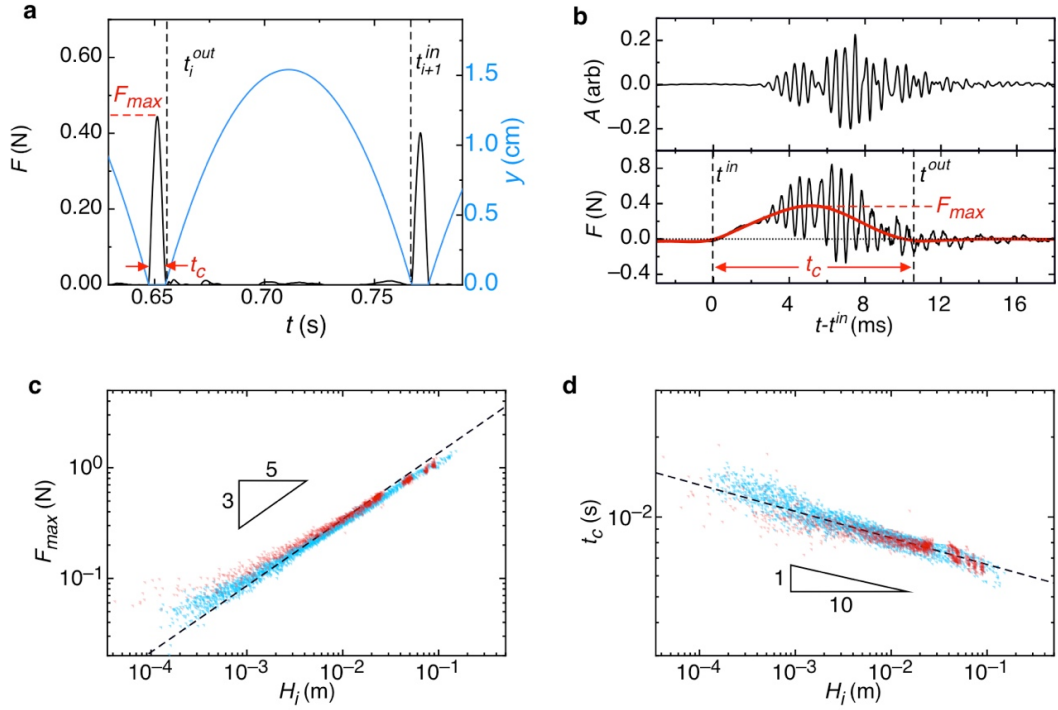

Supplementary Figure 3: Trajectory and impact analysis. **a** Impact force (left y-axis) and calculated vertical trajectory (right y-axis) *vs.* time for a sphere bouncing on a cold surface. We use the force curve to determine the contact intervals and then Eq. 2 to calculate the vertical trajectory in the time between. **b** The audio (top) and force (bottom) curves *vs.* time for a sphere dropped onto a hot surface oscillate as a result of the gap oscillations, but are also affected by mechanical resonances in our setup ranging from about 1-10 kHz. By performing a low pass (<1 kHz) filter on the force data, we are able to clearly identify $t_c$ and $F_{max}$ and analyse the curve in the same way as the cold impact data of panel **a**. The maximum impact force **c** and total contact time **d** for the cold data (blue points) and hot data (red points) both scale with the drop height as predicted by Hertzian theory (black dashed lines, which are calculated with no free parameters from Eqs. 4 and 5, respectively). This verifies our assumption that the cold and hot impacts are largely identical, *i.e.,* there is a separation of scales between the global features of the Hertzian impact and the gap oscillations.

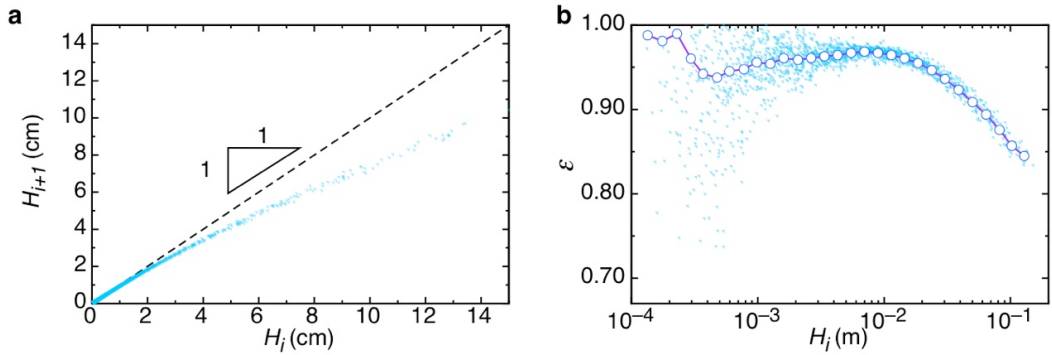

Supplementary Figure 4: Interpolated coefficient of restitution curve. **a** Experimentally determined rebound curve showing $H_{i+1}$ vs. $H_i$ for spheres bouncing on the cold surface. As can be seen, the points deviate significantly from a line, especially at large $H_i$. **b** We see the behaviour more clearly by plotting the coefficient of restitution as a function of drop height on a semi-log scale. The blue dots are individual data points, while the open blue circles are the result of logarithmically binning and averaging. Using the binned points, we create the interpolated curve (purple line) for the coefficient of restitution, $\varepsilon(H_i)$, and calculate the energy injected as described in the Supplementary Text. The uncertainty for a single energy injection measurement (error bars for inset to Fig. 2b) are typically dominated by the spread around the $\varepsilon(H_i)$ curve (especially at lower drop heights).

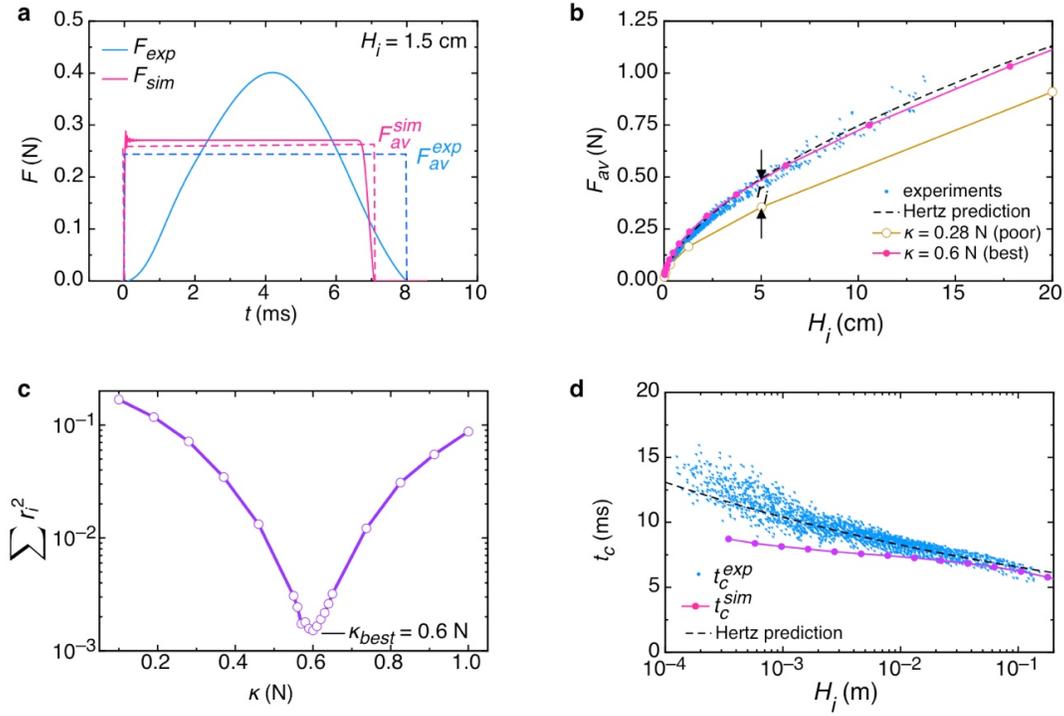

Supplementary Figure 5: Chain spring constant calibration. **a** The force *vs.* time curve of a sphere on a cold surface, (blue solid line) smoothly rises to a maximum and returns to zero, while the same curve for a simulated spring-mass chain resembles a step function. For an appropriate choice of $\kappa$, the average forces will be similar. **b** Average impact force *vs.* $H_i$ for experiments (blue dots), from Hertzian theory (black dashed line), for a poor choice for $\kappa$ (beige open circles) and the best choice of $\kappa$ (closed pink circles). A residual between a numerical point in the poor curve and the theoretical prediction is indicated. **c** Sum of the squared residuals (for curves of $F_{av}$ *vs.* $H_i$) *vs.* $\kappa$, which shows that the best value is $\kappa \approx 0.6$. **d** Matching the average force of experiments and simulations guarantees that the contact times (blue dots for experiments, black dashed line for the Hertz's prediction, and pink dots for simulations) are similar.

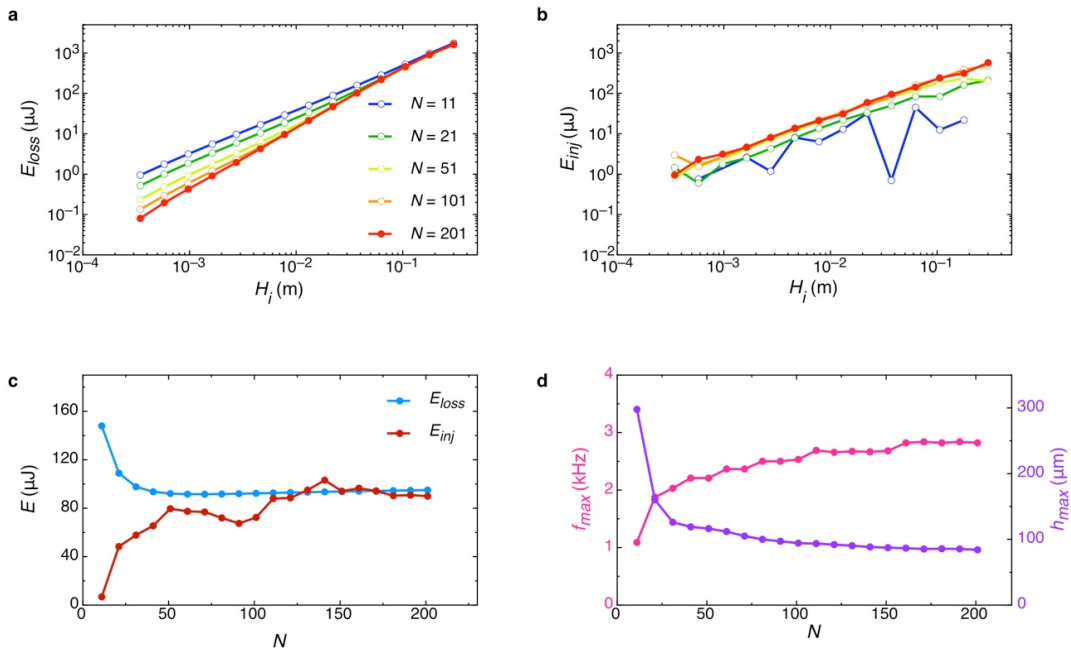

Supplementary Figure 6: Model discretization dependence. These plots demonstrate that our numerical model starts to become independent of the discretization when $N$ is greater than about 50. **a** Plots of the energy lost *vs.* drop height for various $N$ during impact of the spring mass chain into a "cold" surface ($\alpha=0$, all other parameters as in the Supplementary Text). **b** Plots of the energy injected *vs.* drop height for various $N$ (colors same as **a**). **c** Energy lost (blue) and energy injected (red) *vs.* $N$ for a fixed drop height $H_i$ = 3.5 cm. **d** The peak frequency (pink, left axis) and maximum gap oscillation amplitude (purple, right axis) *vs.* $N$ also stabilize beyond $N=50$.

**Descriptions of the Supplementary Videos**

Supplementary Video 1

**Part 1:** Top view of five dyed hydrogel spheres dropped onto a hot (215 ºC) ceramic coated aluminium surface. The spheres frenetically travel around the surface and emit high-pitched screeching noises. The playback is in real time.

**Part 2:** Shallow-angled side view of five dyed hydrogel spheres dropped onto a hot (215 ºC) ceramic coated aluminium surface, which makes it clear that the horizontal motion is mediated by vertical bouncing. This video also illustrates a sphere's sudden death, which occurs as a result of fracture and typically after 2-3 minutes of bouncing. Once inactive on the surface, the polymer in the gel begins to melt and the sphere sticks. The playback is in real time.

Supplementary Video 2

**Part 1:** Side view video of a hydrogel sphere dropped onto a "cold" (25 ºC), gently curved aluminium surface from an initial height of ~6 cm. The sphere loses energy after each impact and quickly comes to rest. The playback is slowed down by a factor of 6.66 and the real time duration is 2.33 s.

**Part 2:** Side view video of a hydrogel sphere dropped onto a "hot" (215 ºC), gently curved aluminium surface from an initial height of ~6 cm. The sphere loses some energy in the initial few impacts, but then reaches a steady bounce height of a few centimeters. The playback is slowed down by a factor of 6.66 and the real time duration is 4.07 s.

**Part 3:** Side view video of a hydrogel sphere dropped onto a hot (215 ºC), gently curved aluminium surface from an initial height of ~2 mm. Despite some variability, the sphere's bounce height increases until it reaches a steady bounce height of a few centimeters. The increase in bounce heights provides clear evidence that the sphere gains energy during impacts. The playback is slowed down by a factor of 6.66 and the real time duration is 1.94 s.

Supplementary Video 3

**Part 1:** Side view, high-speed (5000 fps) video of a single impact onto a cold (25 ºC) ceramic coated aluminium surface from a drop height of ~3.5 cm. Two observations are important: (1) nothing happens at the interface and (2) large wavelength ($\lambda \sim R$) spheroidal modes are excited that serve as the primary source of dissipation during impact. The playback is slowed down by a factor of 333 and the real time duration is 21.8 ms.

**Part 2:** Side view, high-speed (5000 fps) video of a single impact onto a hot (215 ºC) ceramic coated aluminium surface from a drop height of ~3.5 cm. In stark contrast to the previous video, here we see a minute gap that repeatedly opens and closes at the interface. This excites short wavelength (<1 mm) Rayleigh waves which travel upwards around the surface. The large length scale (>1 mm) and low frequency (<1 kHz) features of this impact (*e.g.,* maximum compression, total "contact" time, and the $\lambda \sim R$ spheroidal modes) are essentially the same as in the previous video. The playback is slowed down by a factor of 333 and the real time duration is 20 ms.

Supplementary Video 4

**Part 1:** Side view, high-speed (15625 fps) and zoomed-in video of a single impact onto a hot (215 ºC) ceramic coated aluminium surface from a drop height of ~3.5 cm. The increased spatial and temporal resolution reveals that the gap opens up to heights on the order of ~100 μm at frequencies of 2-3 kHz. These oscillations are the source of the high-pitched screeching noises. The playback is slowed down by a factor of 1042 and the real time duration is 9.6 ms.

Supplementary Video 5

Part 1: Rendered "zoomed in" video of a single spring-mass chain impact onto a "hot" surface from a drop height of 3.5 cm, which shows that the model reproduces the gap oscillations at the interface. The locations of the masses are indicated by dots, and the sphere is rendered by drawing a disk around the instantaneous center of mass of the chain and cutting off the section below the lowest mass. The flat bottom therefore represents the growing contact area used in the numerics. The simulation parameters are those laid out in the Supplementary Text. The video is effectively slowed down by a factor of 1042 and the real time duration is 7.8 ms.

Part 2: Rendered video of a spring-mass chain bouncing on a "cold" surface from an initial height of 6 cm. The chain loses energy to vibrational modes during each impact and approaches rest. We only simulate one bounce at a time and stitch these together by calculating the parabolic flight between bounces. The sphere is rendered in the same way as the previous video, but without dots for the mass locations. The playback is slowed down by a factor of 6.66 and the real time duration is 4.5 s.

Part 3: Rendered video of a spring-mass chain bouncing on a "hot" surface from an initial height of 6 cm, which illustrates the model reproduces the persistent bouncing motion with a steady bounce height of a few centimeters. The playback is slowed down by a factor of 6.66 and the real time duration is 4.5 s.

Part 4: Rendered video of a spring-mass chain dropped onto a "hot" surface from an initial height of 2 mm, which shows that it climbs to the steady-bounce height as in the experiments. The playback is slowed down by a factor of 6.66 and the real time duration is 4.5 s.